\documentclass[aps,prd,twocolumn,showpacs,amsmath,amssymb,floatfix,superscriptaddress,showpacs]{revtex4}

\usepackage{graphicx}
\usepackage{dcolumn}
\usepackage{bm}

\newcommand{\figurewidth}{0.98\hsize}   

\def\met{\mathbin{E\mkern - 11mu/_T}}
\newcommand{\lw}[1]{\smash{\lower1.8ex\hbox{#1}}}

\begin{document}

\preprint{CDF/PUB/EXOTIC/CDFR/9347}
\preprint{Draft Version 3.01}

\title{Search for new particles decaying into dijets
  in proton-antiproton collisions at $\sqrt{s}=1.96$ TeV}

\affiliation{Institute of Physics, Academia Sinica, Taipei, Taiwan 11529, Republic of China} 
\affiliation{Argonne National Laboratory, Argonne, Illinois 60439} 
\affiliation{University of Athens, 157 71 Athens, Greece} 
\affiliation{Institut de Fisica d'Altes Energies, Universitat Autonoma de Barcelona, E-08193, Bellaterra (Barcelona), Spain} 
\affiliation{Baylor University, Waco, Texas  76798} 
\affiliation{Istituto Nazionale di Fisica Nucleare Bologna, $^v$University of Bologna, I-40127 Bologna, Italy} 
\affiliation{Brandeis University, Waltham, Massachusetts 02254} 
\affiliation{University of California, Davis, Davis, California  95616} 
\affiliation{University of California, Los Angeles, Los Angeles, California  90024} 
\affiliation{University of California, San Diego, La Jolla, California  92093} 
\affiliation{University of California, Santa Barbara, Santa Barbara, California 93106} 
\affiliation{Instituto de Fisica de Cantabria, CSIC-University of Cantabria, 39005 Santander, Spain} 
\affiliation{Carnegie Mellon University, Pittsburgh, PA  15213} 
\affiliation{Enrico Fermi Institute, University of Chicago, Chicago, Illinois 60637}
\affiliation{Comenius University, 842 48 Bratislava, Slovakia; Institute of Experimental Physics, 040 01 Kosice, Slovakia} 
\affiliation{Joint Institute for Nuclear Research, RU-141980 Dubna, Russia} 
\affiliation{Duke University, Durham, North Carolina  27708} 
\affiliation{Fermi National Accelerator Laboratory, Batavia, Illinois 60510} 
\affiliation{University of Florida, Gainesville, Florida  32611} 
\affiliation{Laboratori Nazionali di Frascati, Istituto Nazionale di Fisica Nucleare, I-00044 Frascati, Italy} 
\affiliation{University of Geneva, CH-1211 Geneva 4, Switzerland} 
\affiliation{Glasgow University, Glasgow G12 8QQ, United Kingdom} 
\affiliation{Harvard University, Cambridge, Massachusetts 02138} 
\affiliation{Division of High Energy Physics, Department of Physics, University of Helsinki and Helsinki Institute of Physics, FIN-00014, Helsinki, Finland} 
\affiliation{University of Illinois, Urbana, Illinois 61801} 
\affiliation{The Johns Hopkins University, Baltimore, Maryland 21218} 
\affiliation{Institut f\"{u}r Experimentelle Kernphysik, Universit\"{a}t Karlsruhe, 76128 Karlsruhe, Germany} 
\affiliation{Center for High Energy Physics: Kyungpook National University, Daegu 702-701, Korea; Seoul National University, Seoul 151-742, Korea; Sungkyunkwan University, Suwon 440-746, Korea; Korea Institute of Science and Technology Information, Daejeon, 305-806, Korea; Chonnam National University, Gwangju, 500-757, Korea} 
\affiliation{Ernest Orlando Lawrence Berkeley National Laboratory, Berkeley, California 94720} 
\affiliation{University of Liverpool, Liverpool L69 7ZE, United Kingdom} 
\affiliation{University College London, London WC1E 6BT, United Kingdom} 
\affiliation{Centro de Investigaciones Energeticas Medioambientales y Tecnologicas, E-28040 Madrid, Spain} 
\affiliation{Massachusetts Institute of Technology, Cambridge, Massachusetts  02139} 
\affiliation{Institute of Particle Physics: McGill University, Montr\'{e}al, Qu\'{e}bec, Canada H3A~2T8; Simon Fraser University, Burnaby, British Columbia, Canada V5A~1S6; University of Toronto, Toronto, Ontario, Canada M5S~1A7; and TRIUMF, Vancouver, British Columbia, Canada V6T~2A3} 
\affiliation{University of Michigan, Ann Arbor, Michigan 48109} 
\affiliation{Michigan State University, East Lansing, Michigan  48824}
\affiliation{Institution for Theoretical and Experimental Physics, ITEP, Moscow 117259, Russia} 
\affiliation{University of New Mexico, Albuquerque, New Mexico 87131} 
\affiliation{Northwestern University, Evanston, Illinois  60208} 
\affiliation{The Ohio State University, Columbus, Ohio  43210} 
\affiliation{Okayama University, Okayama 700-8530, Japan} 
\affiliation{Osaka City University, Osaka 588, Japan} 
\affiliation{University of Oxford, Oxford OX1 3RH, United Kingdom} 
\affiliation{Istituto Nazionale di Fisica Nucleare, Sezione di Padova-Trento, $^w$University of Padova, I-35131 Padova, Italy} 
\affiliation{LPNHE, Universite Pierre et Marie Curie/IN2P3-CNRS, UMR7585, Paris, F-75252 France} 
\affiliation{University of Pennsylvania, Philadelphia, Pennsylvania 19104}
\affiliation{Istituto Nazionale di Fisica Nucleare Pisa, $^x$University of Pisa, $^y$University of Siena and $^z$Scuola Normale Superiore, I-56127 Pisa, Italy} 
\affiliation{University of Pittsburgh, Pittsburgh, Pennsylvania 15260} 
\affiliation{Purdue University, West Lafayette, Indiana 47907} 
\affiliation{University of Rochester, Rochester, New York 14627} 
\affiliation{The Rockefeller University, New York, New York 10021} 
\affiliation{Istituto Nazionale di Fisica Nucleare, Sezione di Roma 1, $^{aa}$Sapienza Universit\`{a} di Roma, I-00185 Roma, Italy} 

\affiliation{Rutgers University, Piscataway, New Jersey 08855} 
\affiliation{Texas A\&M University, College Station, Texas 77843} 
\affiliation{Istituto Nazionale di Fisica Nucleare Trieste/Udine, I-34100 Trieste, $^{bb}$University of Trieste/Udine, I-33100 Udine, Italy} 
\affiliation{University of Tsukuba, Tsukuba, Ibaraki 305, Japan} 
\affiliation{Tufts University, Medford, Massachusetts 02155} 
\affiliation{Waseda University, Tokyo 169, Japan} 
\affiliation{Wayne State University, Detroit, Michigan  48201} 
\affiliation{University of Wisconsin, Madison, Wisconsin 53706} 
\affiliation{Yale University, New Haven, Connecticut 06520} 
\author{T.~Aaltonen}
\affiliation{Division of High Energy Physics, Department of Physics, University of Helsinki and Helsinki Institute of Physics, FIN-00014, Helsinki, Finland}
\author{J.~Adelman}
\affiliation{Enrico Fermi Institute, University of Chicago, Chicago, Illinois 60637}
\author{T.~Akimoto}
\affiliation{University of Tsukuba, Tsukuba, Ibaraki 305, Japan}
\author{B.~\'{A}lvarez~Gonz\'{a}lez$^q$}
\affiliation{Instituto de Fisica de Cantabria, CSIC-University of Cantabria, 39005 Santander, Spain}
\author{S.~Amerio$^w$}
\affiliation{Istituto Nazionale di Fisica Nucleare, Sezione di Padova-Trento, $^w$University of Padova, I-35131 Padova, Italy} 

\author{D.~Amidei}
\affiliation{University of Michigan, Ann Arbor, Michigan 48109}
\author{A.~Anastassov}
\affiliation{Northwestern University, Evanston, Illinois  60208}
\author{A.~Annovi}
\affiliation{Laboratori Nazionali di Frascati, Istituto Nazionale di Fisica Nucleare, I-00044 Frascati, Italy}
\author{J.~Antos}
\affiliation{Comenius University, 842 48 Bratislava, Slovakia; Institute of Experimental Physics, 040 01 Kosice, Slovakia}
\author{G.~Apollinari}
\affiliation{Fermi National Accelerator Laboratory, Batavia, Illinois 60510}
% ...

\collaboration{CDF Collaboration\footnote{With visitors from $^a$University of Massachusetts Amherst, Amherst, Massachusetts 01003,
$^b$Universiteit Antwerpen, B-2610 Antwerp, Belgium, 
$^c$University of Bristol, Bristol BS8 1TL, United Kingdom,
$^d$Chinese Academy of Sciences, Beijing 100864, China, 
$^e$Istituto Nazionale di Fisica Nucleare, Sezione di Cagliari, 09042 Monserrato (Cagliari), Italy,
$^f$University of California Irvine, Irvine, CA  92697, 
$^g$University of California Santa Cruz, Santa Cruz, CA  95064, 
$^h$Cornell University, Ithaca, NY  14853, 
$^i$University of Cyprus, Nicosia CY-1678, Cyprus, 
$^j$University College Dublin, Dublin 4, Ireland,
$^k$University of Edinburgh, Edinburgh EH9 3JZ, United Kingdom, 
$^l$Universidad Iberoamericana, Mexico D.F., Mexico,
$^m$Queen Mary, University of London, London, E1 4NS, England,
$^n$University of Manchester, Manchester M13 9PL, England, 
$^o$Nagasaki Institute of Applied Science, Nagasaki, Japan, 
$^p$University of Notre Dame, Notre Dame, IN 46556,
$^q$University de Oviedo, E-33007 Oviedo, Spain, 
$^r$Texas Tech University, Lubbock, TX  79409, 
$^s$IFIC(CSIC-Universitat de Valencia), 46071 Valencia, Spain,
$^t$University of Virginia, Charlottesville, VA  22904,
$^u$Bergische Universit\"at Wuppertal, 42097 Wuppertal, Germany,
$^{cc}$On leave from J.~Stefan Institute, Ljubljana, Slovenia, 
}}
\noaffiliation


\date{\today}

\begin{abstract}
We present a search for new particles whose decays produce two jets
(dijets) using proton-antiproton collision data
corresponding to an integrated luminosity of 1.13 fb$^{-1}$
collected with the CDF II detector.
The measured dijet mass spectrum is found to be consistent with
next-to-leading-order perturbative QCD predictions, and 
no significant evidence of new particles is found.
We set upper limits at the 95\% confidence level on 
cross sections times the branching fraction 
for the production of new particles decaying into
dijets with both jets having a rapidity magnitude $|y|<1$.
These limits are used to determine the mass exclusions for the
excited quark, axigluon, flavor-universal coloron, $E_6$ diquark,
color-octet techni-$\rho$, $W'$, and $Z'$.
\end{abstract}


\maketitle

%
\section{\label{sec:Intro}Introduction}

Within the standard model (SM), two-jet (dijet) events are 
produced in proton-antiproton ($p\bar p$) collisions predominantly
from hard
quantum chromodynamics (QCD) interactions of two partons.
The fragmentation and hadronization of the outgoing partons
produce hadronic jets.
The dijet mass spectrum predicted by QCD falls smoothly and
steeply with increasing dijet mass.
Many extensions of the SM predict the existence of new massive
particles
that decay into two energetic partons (quarks, $q$, or gluons, $g$),
which can potentially be observed as 
a narrow resonance in the dijet mass spectrum.
Such particles include the
excited quark~\cite{exq}, 
axigluon~\cite{axigluon}, 
flavor-universal coloron~\cite{coloron}, 
color-octet techni-$\rho$~\cite{technirho}, 
Randall-Sundrum (RS) graviton~\cite{rsg},
$W'$, 
$Z'$~\cite{wzprime}, 
and diquark in the string-inspired $E_6$ model~\cite{E6}.

Here we briefly discuss the theoretical models for these new particles.
In the SM, the quarks are considered as fundamental particles.
However, the presence of their generational structure 
and mass hierarchy motivate models of quark
compositeness in which the quarks consist of more fundamental
particles.
If a quark is a composite particle, an excited state of a quark $q^*$
is expected, which decays to $qg$~\cite{exq}.
In chiral color models,
the $SU(3)$ gauge group of QCD results from the spontaneous breaking of 
the chiral color gauge group of $SU(3)\times SU(3)$.
Any model of chiral color predicts the presence of the axigluon, 
a massive axial vector gluon, that decays to $q\bar q$~\cite{axigluon}.
The flavor-universal coloron model also embeds the
$SU(3)$ of QCD in a larger gauge group and predicts the presence of a
color-octet coloron which decays to $q\bar q$~\cite{coloron}.
Technicolor models seek to explain electroweak symmetry breaking
via the dynamics of new interactions among techniquarks.
The models of extended technicolor and topcolor-assisted technicolor 
predict the presence of a color-octet techni-$\rho$
($\rho_{T8}$) which decays to $q\bar q$ or $gg$~\cite{technirho}.
The RS model of a warped extra dimension offers a solution
for the hierarchy between the electroweak scale and Planck scale
$\bar M_{Pl}$ by introducing an extra spacial dimension~\cite{rsg}.
This model predicts a Kaluza-Klein tower of graviton states (RS
gravitons) which decay to $q\bar q$ or $gg$.
The grand unified theories (GUT) based on 
larger gauge groups, e.g., $E_6$ and $SO(10)$, or
left-right-symmetric models~\cite{left-right-symmetry}
often introduce additional gauge bosons, such as $W'$ and $Z'$, which
decay to $q\bar q'$ and $q\bar q$, respectively~\cite{wzprime}.
The $E_6$ GUT model also predicts the presence of a diquark which decays
to $qq$ or $\bar q \bar q$~\cite{E6}.

In the past, the UA2~\cite{ua2}, CDF~\cite{cdf1a,cdf1b} and
D0~\cite{dzero} experiments searched for resonances in the
dijet mass spectrum and set limits on their production.
In this article, we present a first measurement of the dijet mass
spectrum and a search for massive particles which decay into dijets 
in $p\bar p$ collisions at the center-of-mass energy
$\sqrt{s}=1.96$ TeV.
This analysis uses data corresponding to an integrated luminosity of
1.13~${\rm fb}^{-1}$
collected between February 2002 and February 2006 with the
CDF II detector at the Fermilab Tevatron.

The measurement of the dijet mass spectrum is also an important
test of perturbative QCD (pQCD) predictions.
It provides complementary information to the inclusive jet cross
section
measurements~\cite{KT_PRL,Midpoint_RC,KT_PRD,Dzero_Run2_incjet,Midpoint_PRD},
and comparisons
of the measurement with pQCD predictions provide constraints on the
parton distribution functions (PDFs) of the proton, in
particular at high momentum fraction $x$ ($x \gtrsim 0.3$) where
the gluon distribution is not well constrained~\cite{CTEQ6.1M}.

%
\section{\label{sec:detector}The CDF Detector}

The CDF II detector 
is described in detail elsewhere~\cite{CDF_detect_A}.
Here, the components that are relevant to this search are
briefly described.
Surrounding the beam pipe, there is a tracking system consisting of
a silicon microstrip detector, a cylindrical drift chamber, and a
solenoid magnet that provides a 1.4 T magnetic field.
The central and plug calorimeters, 
which cover the pseudorapidity regions of
$|\eta|<1.1$ and $1.1<|\eta|<3.6$~\cite{coordinate},
respectively,
surround the tracking system with a projective tower geometry and
measure the energy of interacting particles.
The calorimeters are segmented into electromagnetic and hadronic
sections that consist of lead-scintillator and iron-scintillator,
respectively.
In the central region, the
calorimeter consists of 48 modules, segmented into towers of
granularity
$\Delta \eta \times \Delta\phi \approx  0.1 \times 0.26$.
The energy resolution of the central electromagnetic calorimeter for
electrons is
$\sigma(E_T) / E_T = 13.5\%/\sqrt{E_T ({\rm GeV})} \oplus 1.5\%$,
while the energy resolution of the central hadron calorimeter for
charged pions that do not interact in the electromagnetic section is
$\sigma(E_T) / E_T = 50\%/\sqrt{E_T ({\rm GeV})} \oplus 3\%$,
where $E_T$ is the transverse energy~\cite{coordinate}.
The wall hadron calorimeter covers the gap in the projective tower
geometry between the central and plug hadron calorimeters,
corresponding to $0.7<|\eta|<1.3$, with segmentation similar to that
of the central calorimeter.
The energy resolution of the wall hadron calorimeter is
$\sigma(E_T)/E_T = 75\%/\sqrt{E_T({\rm GeV})} \oplus 4\%$
for charged pions that do not interact in the electromagnetic section.
A system of Cherenkov counters,
located around the beam pipe and inside the plug
calorimeters, is used to
measure the number of inelastic $p\bar p$ collisions per bunch
crossing and thereby the luminosity.

%
\section{\label{sec:data_evtsel}Datasets and event
  selection}

The dijet data used in this search were collected using a three-level
online event selection (trigger) system, and are identical to the data
used in Ref.~\cite{Midpoint_PRD}.
The four trigger selections used in this analysis are referred to as
``jet20'', ``jet50'', ``jet70'', and ``jet100'' according to the
$E_T$ threshold (in GeV) of calorimeter clusters
reconstructed using a cone algorithm~\cite{incjet_run1}
with cone radius $R_{cone}=0.7$.
The jet20, jet50, and jet70 trigger rates are 
randomly reduced (prescaled) to avoid saturating the
bandwidth of the data acquisition system.
After prescaling, these trigger datasets
correspond to 1.44, 32.5, and 143 pb$^{-1}$ of integrated luminosity,
respectively.
The jet100 trigger is not prescaled.

Jets are reconstructed from the energy depositions 
in the calorimeter towers with the transverse momentum
$p_T$~\cite{coordinate} above $0.1$ GeV/$c$.
Jets are formed from the four-vectors of calorimeter towers~\cite{caltow} 
using the cone-based midpoint jet clustering
algorithm~\cite{CDF_midpoint1,CDF_midpoint2,Midpoint_RC} with cone radius
$R_{cone}=0.7$.
The kinematics of a jet is defined by the four-vector recombination
scheme~\cite{CDF_midpoint1}.

Cosmic ray and beam loss background events are removed by
requiring 
$\met/ \sqrt{\sum E_{T}} < \min(3+0.0125\times
p_{T}^{jet1},6)$,
where $\met$ and $\sum E_T$ are the missing $E_T$ and sum
$E_T$~\cite{missingEt}, respectively, 
and $p_{T}^{jet1}$
(GeV/{\it c}) is the $p_T$ of the leading jet in the event before the
corrections described in Sec.~\ref{sec:jetcor_mjj_spectrum} are applied.
This requirement makes the fraction of background events negligible,
and has an efficiency
of $\gtrsim95$\% for dijet events.
In order to ensure good coverage of each event by the detectors,
the primary event vertex is required to be within 60 cm of the
center of the detector along the $z$-axis~\cite{coordinate}.
The efficiency for this requirement is determined to be 96\% from the
distribution of primary event vertices along $z$
measured using a sample of minimum bias events.

%
\section{\label{sec:jetcor_mjj_spectrum}Jet corrections and
  measurement of the dijet mass spectrum}

The jet energies measured by the calorimeters are affected by
instrumental effects such as calorimeter non-uniformity,
non-linearity, and energy smearing.
We correct for these biases in several steps~\cite{JESNIM}.
First, an $\eta$-dependent relative correction is applied
to equalize the response of the calorimeter.
The equalized jet $p_T$ is then corrected for the effects of pileup,
i.e., additional $p\bar p$ interactions in the same bunch
crossing.
The pileup correction subtracts $0.97\pm0.29$ GeV/{\it c} for each
additional primary vertex from the measured jet $p_T$.
Then, a $p_T$-dependent correction is applied to
account for, on average, the under-measured hadron energy due to
the non-linearity of the calorimeter response.
The correction factors are 1.19 and 1.06
at jet $p_T=90$ and 600 GeV/$c$, respectively.
After these corrections we 
reconstruct the dijet mass $m_{jj}$ from the four-vectors of the two
highest corrected-$p_T$ jets.
We form the dijet mass spectrum as:
\begin{equation}
\left.
\frac{d\sigma}{dm_{jj}}\right|_i = 
  \frac{1}{{\cal L}_i\cdot\epsilon_{trig,i}}
  \frac{n_{i}}{\Delta m_{jj,i}},
  \label{eqn:dsigma_dmjj}
\end{equation}
where $n_{i}$ is the observed number of events,
$\epsilon_{trig,i}$ is the trigger efficiency, 
$\Delta m_{jj,i}$ is the bin width,
and ${\cal L}_i$ is the integrated luminosity
of the trigger dataset used for the $i$-th
dijet mass bin.
The bin width is set to 10\% of the dijet mass
which approximately corresponds to the dijet mass resolution.
We count only events 
in which both of the leading two jets have a rapidity magnitude
$|y|$~\cite{coordinate} less than 1.
We use the jet20, jet50, jet70, and jet100 datasets
for the dijet mass regions of
180--241, 241--321, 321--427, and above 427 $\mbox{GeV}/c^2$,
respectively, where the trigger efficiencies are higher than 99.8\%.
This ensures a negligible uncertainty from trigger efficiency
measurements.

In order to compare the measured dijet mass spectrum 
with QCD predictions 
at the hadron level~\cite{Midpoint_PRD},
the measured spectrum must be further corrected for the bin-by-bin
migration effect due to the finite resolution of the $m_{jj}$ measurement
and the efficiencies of the offline event selection requirements.
We obtain this correction using QCD dijet events generated by
the {\sc pythia} 6.2~\cite{pythia} Monte Carlo simulation program, that
have passed through the CDF detector simulation~\cite{cdfsim}.
The {\sc pythia} events are generated with Tune A~\cite{TuneA},
which refers to the set of parameters describing multiple-parton
interactions and initial state radiation
that have been tuned to reproduce the energy observed in the
region transverse to the leading jet~\cite{run1ue}.
It has also been shown to provide a reasonable description of the
measured energy distribution inside a jet~\cite{jetshape}.
The correction is determined on a bin-by-bin basis by taking
the ratio of a hadron-level cross section to a calorimeter-level
cross section.
The hadron-level cross section is defined using hadron-level jets
clustered from the final state stable
particles~\cite{stable_particles} in {\sc pythia} with
the same jet clustering algorithm as the one used to cluster
calorimeter towers. The leading two jets are required to have
$|y|<1$.
The calorimeter-level cross section is obtained by analyzing the
{\sc pythia} events using the same analysis chain as for the data.
Since the correction depends on the dijet mass spectrum, the {\sc
  pythia} events are reweighted to match the dijet mass spectrum
measured in data before the correction factor is calculated.
The size of the correction ranges from $\sim1.2$ at low $m_{jj}$
($200$ GeV/$c^2$) to $\sim1.5$ at high $m_{jj}$ ($1250$ GeV/$c^2$).

\begin{figure}[tp]\centering\leavevmode
  \includegraphics[width=\figurewidth,bb=0 45 480 670,clip=]
  {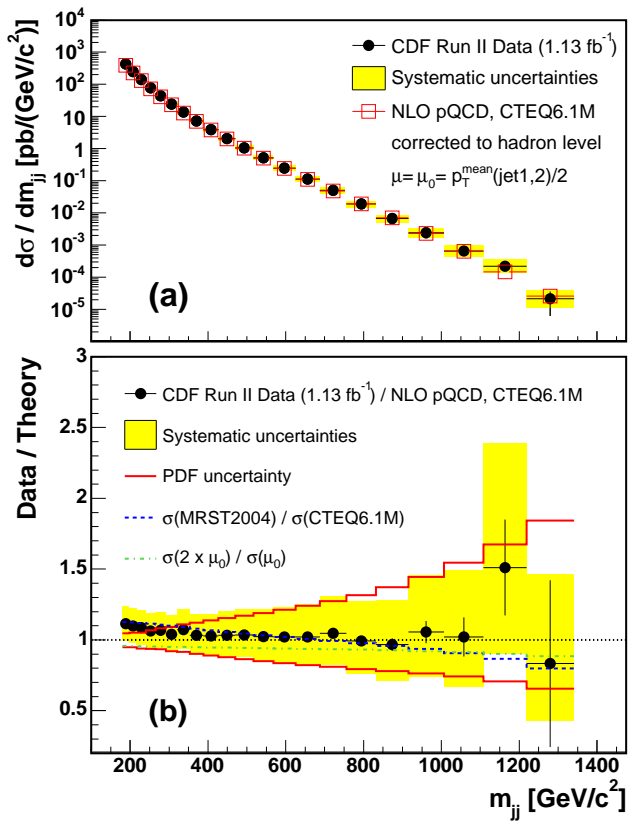}
  \caption{(a) The measured dijet mass spectrum
    for both jets to have $|y|<1$
    compared to the NLO pQCD prediction obtained using the CTEQ6.1 PDFs. 
    (b) The ratio of the data to the NLO pQCD prediction.
    The experimental systematic uncertainties, theoretical uncertainties
    from PDF, the ratio of MRST2004/CTEQ6.1, and
    the dependence on the choice of renormalization and factorization
    scales 
    are also shown.
    An additional 6\% uncertainty in the determination of the luminosity
    is not shown.}
  \label{fig:data_nlo}
\end{figure}

The systematic uncertainties arise mainly from four sources:
the jet energy scale,
the jet energy resolution,
the unfolding correction,
and the integrated luminosity.
The dominant source 
is from the absolute jet energy scale.
The size of the uncertainty in the cross section varies from
10\% at low $m_{jj}$ to $^{+74}_{-47}$\% at high $m_{jj}$.
The uncertainty in the relative jet energy scale introduces a
$3$\% uncertainty on the measured cross section at low $m_{jj}$ and
$^{+10}_{-9}$\% at high $m_{jj}$.
The uncertainty in the modeling of jet energy smearing
is estimated from the difference between the
data and {\sc pythia} samples using the bisector
method~\cite{Bisector}, and
it introduces an uncertainty in the cross section of 1\% at low
$m_{jj}$ and $^{+6}_{-5}$\% at high $m_{jj}$.
The difference between the unfolding correction from a {\sc pythia}
sample and that from a sample generated by the {\sc herwig}
6.5~\cite{HERWIG} Monte Carlo simulation program is also taken as a systematic
uncertainty to account for the
uncertainty in the modeling of jet fragmentation.
This uncertainty is $2$\% at low $m_{jj}$ and $8$\% at high $m_{jj}$.
The uncertainty in the determination of the integrated luminosity is 6\%,
independent of $m_{jj}$.
The total uncertainty is $^{+13}_{-12}$\% at low $m_{jj}$ and
$^{+76}_{-49}$\% at high $m_{jj}$.
The measured dijet mass spectrum after all the corrections discussed
above are applied is shown together with the
statistical and systematic uncertainties in Fig.~\ref{fig:data_nlo}(a).

\section{Comparisons with QCD predictions}

The measured dijet mass spectrum is compared
in Fig.~\ref{fig:data_nlo}
to the next-to-leading-order
perturbative QCD (NLO pQCD) predictions from fastNLO~\cite{FastNLO}.
The predictions were obtained using 
the CTEQ6.1~\cite{CTEQ6.1M} PDFs 
with the renormalization and factorization scales both set to $\mu_0$,
the average $p_T$ of the leading two jets.
Jets are reconstructed by the midpoint algorithm with cone radius
$R_{cone}=0.7$.
The maximum separation between two partons
merged into a jet is set to $R_{cone}\times R_{sep}$~\cite{Rsep},
and $R_{sep}$ is set to 1.3~\cite{incjet_run1,CDF_midpoint2}.
Setting the renormalization and factorization scales to $2\mu_0$
instead of $\mu_0$ reduces the cross section prediction by 5--10\%,
and setting $R_{sep}=2$ increases the cross section by $\lesssim10$\%.
The PDF uncertainties estimated from 40 CTEQ6.1 error PDFs and
the ratio of the predictions using MRST2004~\cite{MRST2004} and
CTEQ6.1 are shown in Fig.~\ref{fig:data_nlo}(b).
The PDF uncertainty is the dominant theoretical uncertainty for most
of the $m_{jj}$ range.
The NLO pQCD predictions for jets clustered from partons need to be
corrected for non-perturbative underlying event and hadronization
effects.
The multiplicative parton-to-hadron-level correction ($C_{p\to h}$) 
is determined on a bin-by-bin basis from a ratio of two dijet mass
spectra.
The numerator is the nominal hadron-level dijet mass spectrum
from the {\sc pythia} Tune A samples,
and the denominator is the dijet mass spectrum 
obtained from jets formed from partons before hadronization
in a sample simulated with an underlying event turned off.
We assign the
difference between the corrections obtained using {\sc herwig} and 
{\sc pythia} Tune A as 
the uncertainty on the $C_{p\to h}$ correction.
The $C_{p\to h}$ correction is $1.16\pm0.08$ at
low $m_{jj}$ and $1.02\pm0.02$ at high $m_{jj}$.
Figure~\ref{fig:data_nlo} shows the ratio of the measured spectrum
to the NLO pQCD predictions corrected for the non-perturbative effects.
The data and theoretical predictions are found to be in good
agreement.
To quantify the agreement, we performed a $\chi^2$ test which is
the same as the one used in the inclusive jet cross section
measurements~\cite{KT_PRD,Midpoint_PRD}.
The test treats the systematic uncertainties from different sources and
uncertainties on $C_{p\to h}$ as independent but fully correlated over
all $m_{jj}$ bins and yields $\chi^2/{\rm n.d.f.}=21/21$.

%
\section{\label{sec:resonance_research}Search for dijet mass resonances}

\begin{figure}[tp]\centering\leavevmode
  \includegraphics[width=\figurewidth,bb=0 145 475 632,clip=]
  {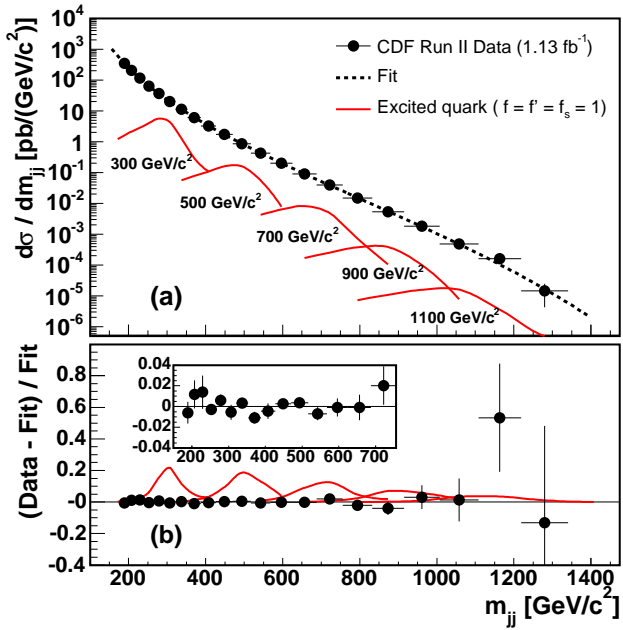}
  \caption{(a) The measured dijet mass spectrum (points) fitted to
    Eq.~(\ref{eqn:ab}) (dashed curve).
    The bin-by-bin unfolding corrections is not applied.
    Also shown are the predictions from the excited quark, $q^*$,
    simulations for
    masses of
    300, 500, 700, 900, and 1100 $\mbox{GeV}/c^2$, respectively
    (solid curves).
    (b) The fractional difference between the measured dijet mass
    distribution and the fit (points) compared to the predictions for
    $q^*$ signals divided by the fit to the measured dijet mass spectrum
    (curves).
    The inset shows the expanded view in which 
    the vertical scale is restricted to $\pm 0.04$.
  }
  \label{fig:mjj_data_fit_ab}
\end{figure}

We search for narrow mass resonances in the measured dijet mass
spectrum by fitting the measured spectrum to a smooth functional
form and by looking for data points that show significant excess from
the fit.
We fit the measured dijet mass spectrum before the bin-by-bin
unfolding correction is applied.
We use the following functional form:
\begin{equation}
\frac{d\sigma}{dm_{jj}} = p_0 (1-x)^{p_1} /
x^{p_2+p_3\cdot\ln(x)},\,\,\,\,\,x=m_{jj}/\sqrt{s},
\label{eqn:ab}
\end{equation}
where $p_0$, $p_1$, $p_2$, and $p_3$ are free parameters.
This form fits well the dijet mass spectra from
{\sc pythia}, {\sc herwig}, and NLO pQCD predictions.
The result of the fit to the measured dijet mass spectrum is shown in
Fig.~\ref{fig:mjj_data_fit_ab}.
Equation~(\ref{eqn:ab}) fits the measured dijet mass spectrum 
well with $\chi^2/{\rm n.d.f.}=16/17$.
We find no evidence for the existence of a resonant structure,
and in the next section we use the data to set limits on new particle
production.
%

%
\section{\label{sec:limits}Limits on new particle production}

Several theoretical models which predict the existence of
new particles that decay into dijets are considered
in this search.
For the excited quark $q^*$ which decays to $qg$,
we set its couplings to
the SM $SU(2)$, $U(1)$, and $SU(3)$ gauge groups to be 
$f=f'=f_s=1$~\cite{exq}, 
respectively, and
the compositeness scale to the mass of $q^*$.
For the RS graviton $G^*$ that decays into $q\bar q$ or
$gg$, 
we use the model parameter $k/\bar M_{Pl}=0.1$
which determines the couplings of the graviton to the 
SM particles.
The production cross section increases with increasing 
$k/\bar M_{Pl}$; however, 
values of $k/\bar M_{Pl}\gg 0.1$ are disfavored
theoretically~\cite{rsg2}. 
For $W'$ and $Z'$, which decay to $q\bar{q'}$ and $q\bar q$
respectively, we use the SM couplings.
The leading-order production cross sections of the RS
graviton, $W'$, and $Z'$ are multiplied by a factor of 1.3 to account
for higher-order effects in the strong coupling constant
$\alpha_s$~\cite{kfactor}.
All these models are simulated with {\sc pythia} Tune A.
Signal events of these models from {\sc pythia} are then passed
through the CDF detector simulation.

\begin{figure}[tp]\centering\leavevmode
  \includegraphics[width=\figurewidth,bb=-20 0 587 254]
  {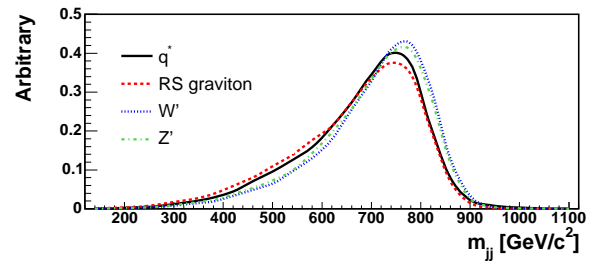}
  \caption{Dijet mass distributions for simulated signals of the 
    $q^*$, RS graviton, $W'$, and $Z'$ with the mass of 800
    GeV/$c^2$.}
  \label{fig:plot_resonance_Cal_800}
\end{figure}

The dijet mass distributions from $q^*$ simulations
with masses 300, 500, 700, 900, and 1100 GeV/$c^2$ are shown
in Fig.~\ref{fig:mjj_data_fit_ab}.
The dijet mass distributions for the $q^*$, RS
graviton, $W'$, and $Z'$ simulations with the mass of 800
$\mbox{GeV}/c^2$ are
shown together in Fig.~\ref{fig:plot_resonance_Cal_800}.
The shapes of the distributions are mainly determined by the jet energy
resolution and QCD radiation which leads to tails on the low mass
side.
Since the natural width of these particles is substantially smaller
than the width from the jet energy resolution, all the dijet mass
distributions appear similar.
However, the dijet mass resonance distributions
are somewhat broader for $q^*$ and
RS gravitons than for $W'$ and $Z'$ because $q^*$
and RS gravitons can decay into the mode containing
gluons, unlike $W'$ and $Z'$. 
Gluons radiate more than quarks and
tend to make the resulting dijet mass distributions broader.
As a result, the cross section limits obtained based on the $q^*$
and RS graviton resonance shapes are about 20\% larger
than those obtained with the $W'$ and $Z'$ resonance shapes.

\begin{figure}[tp]\centering\leavevmode
  \includegraphics[width=\figurewidth,bb=0 0 460 730,clip=]
  {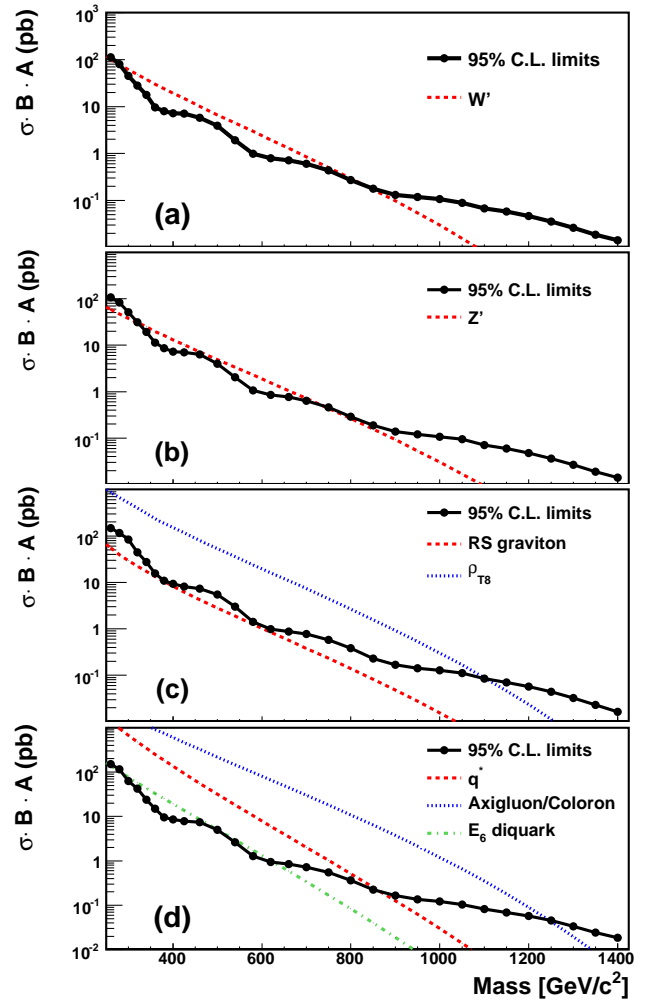}
  \caption{Observed 95\% C.L. upper limits on new particle
    production cross sections times the branching fraction 
    to dijets obtained with the signal shapes from
    (a) $W'$, (b) $Z'$, (c) RS graviton, and (d) $q^*$ production.
    Also shown are the cross section predictions for the production of
    $W'$, $Z'$, 
    RS graviton,
    $\rho_{T8}$,
    $q^*$, 
    axigluon, flavor-universal coloron, and 
    $E_6$ diquark for the set of parameters described in the text.
    The limits and theoretical predictions are for events 
    in which both of the leading two jets have $|y|<1$.}
  \label{fig:limit_theory_vs_mass}
\end{figure}

We also consider production of
the axigluon $A$ that decays into $q\bar q$, 
$E_6$ diquark $D$ $(D^c)$ that decays into $\bar q\bar q$ $(qq)$,
and color-octet techni-$\rho$ ($\rho_{T8}$) that decays into $q\bar
q$ or $gg$. 
Their lowest-order theoretical predictions for 
$\sigma^{sig}\equiv \sigma\cdot\mathcal{B}\cdot\mathcal{A}$
are shown in
Fig.~\ref{fig:limit_theory_vs_mass} along with the predictions for
the other models described above as a function of new particle mass,
where $\sigma$ is the new particle production cross section,
$\mathcal{B}$ is the branching fraction to dijets, 
and $\mathcal{A}$ is the kinematical acceptance for each of the leading two
jets to have $|y|<1$.
In addition, the flavor-universal coloron
$C$ 
which decays to $q\bar q$ is considered.
The cross section for the coloron is always larger than or equal to
that for the axigluon,
so the limits on the axigluon apply to the coloron as well.
For $\rho_{T8}$ production, predictions are for
the mass-degenerate $\rho_{T8}$ with the standard
topcolor-assisted-technicolor couplings and with the set of
parameters in~\cite{technirho_params}.

We set upper limits on $\sigma^{sig}\equiv \sigma\cdot\mathcal{B}\cdot
\mathcal{A}$ as follows.
We use the likelihood function
$L=\prod_{i}\mu_{i}^{n_{i}}\exp(-\mu_{i})/n_{i}!$,
where $\mu_{i}\equiv\sigma^{sig}{\cal
  L}_{i}\epsilon_{i}n_{i}^{sig}/n^{sig}_{tot}+n_{i}^{QCD}$ is the
predicted number of events, $\epsilon_{i}$ is the event
selection efficiency in the $i^{\rm th}$ dijet mass bin,
and $n_{i}^{sig}/n^{sig}_{tot}$ is the predicted fraction of signal
events in bin $i$.
We model the QCD dijet mass spectrum with Eq.~(\ref{eqn:ab}) and use
Eq.~(\ref{eqn:dsigma_dmjj}) to
extract $n_{i}^{QCD}$ from the differential cross section
$d\sigma/dm_{jj}$.
For each value of $\sigma^{sig}$ we maximize the likelihood with
respect to the four parameters in Eq.~(\ref{eqn:ab}).
We integrate this profiled likelihood over Bayesian priors for
the parameters describing the systematic uncertainties~\cite{Bayes},
and we use a flat prior on $\sigma^{sig}$ to extract Bayesian
upper limits on that parameter.
Although this procedure uses Bayesian techniques, we verified that 
the resulting upper limits have good frequentist coverage.

\begingroup
\begin{table}
\caption{Observed 95\% C.L. upper limits on the new particle
    production cross sections times branching fraction 
    to dijets times the acceptance for both jets to have $|y|<1$
    obtained with the
    signal shapes from $W'$, $Z'$, RS graviton ($G^*$), and
    $q^{*}$ production.}
{
\begin{ruledtabular}
\begin{tabular}{cllll}
Mass & \multicolumn{4}{c}{95\% C.L. 
  $\sigma\cdot\mathcal{B}\cdot\mathcal{A}$ [pb]} \\
(GeV/$c^2$) 
 & \multicolumn{1}{c}{$W'$} 
 & \multicolumn{1}{c}{$Z'$} 
 & \multicolumn{1}{c}{$G^*$} 
 & \multicolumn{1}{c}{$q^{*}$} \\ \hline
 260 & $1.1\times10^{ 2}$ & $1.1\times10^{ 2}$ & $1.5\times10^{ 2}$& $1.5\times10^{ 2}$ \\
 280 & $8.1\times10^{ 1}$ & $8.3\times10^{ 1}$ & $1.2\times10^{ 2}$& $1.1\times10^{ 2}$ \\
 300 & $4.5\times10^{ 1}$ & $5.1\times10^{ 1}$ & $8.3\times10^{ 1}$& $6.3\times10^{ 1}$ \\
 320 & $2.8\times10^{ 1}$ & $3.1\times10^{ 1}$ & $4.4\times10^{ 1}$& $4.2\times10^{ 1}$ \\
 340 & $1.8\times10^{ 1}$ & $1.9\times10^{ 1}$ & $2.8\times10^{ 1}$& $2.4\times10^{ 1}$ \\
 360 & $1.0\times10^{ 1}$ & $1.1\times10^{ 1}$ & $1.6\times10^{ 1}$& $1.5\times10^{ 1}$ \\
 380 & $8.0\times10^{ 0}$ & $8.6\times10^{ 0}$ & $1.1\times10^{ 1}$& $1.0\times10^{ 1}$ \\
 400 & $7.2\times10^{ 0}$ & $7.3\times10^{ 0}$ & $9.2\times10^{ 0}$& $8.6\times10^{ 0}$ \\
 425 & $7.1\times10^{ 0}$ & $7.0\times10^{ 0}$ & $8.2\times10^{ 0}$& $7.8\times10^{ 0}$ \\
 460 & $5.8\times10^{ 0}$ & $6.3\times10^{ 0}$ & $7.4\times10^{ 0}$& $7.4\times10^{ 0}$ \\
 500 & $3.9\times10^{ 0}$ & $4.0\times10^{ 0}$ & $5.5\times10^{ 0}$& $5.0\times10^{ 0}$ \\
 540 & $1.9\times10^{ 0}$ & $2.0\times10^{ 0}$ & $3.0\times10^{ 0}$& $2.6\times10^{ 0}$ \\
 580 & $1.0\times10^{ 0}$ & $1.1\times10^{ 0}$ & $1.4\times10^{ 0}$& $1.3\times10^{ 0}$ \\
 620 & $8.0\times10^{-1}$ & $8.5\times10^{-1}$ & $1.0\times10^{ 0}$& $1.0\times10^{ 0}$ \\
 660 & $7.2\times10^{-1}$ & $7.6\times10^{-1}$ & $8.8\times10^{-1}$& $8.4\times10^{-1}$ \\
 700 & $6.0\times10^{-1}$ & $6.4\times10^{-1}$ & $7.8\times10^{-1}$& $7.3\times10^{-1}$ \\
 750 & $4.3\times10^{-1}$ & $4.6\times10^{-1}$ & $5.7\times10^{-1}$& $5.6\times10^{-1}$ \\
 800 & $2.7\times10^{-1}$ & $2.9\times10^{-1}$ & $3.9\times10^{-1}$& $3.7\times10^{-1}$ \\
 850 & $1.8\times10^{-1}$ & $1.9\times10^{-1}$ & $2.3\times10^{-1}$& $2.3\times10^{-1}$ \\
 900 & $1.3\times10^{-1}$ & $1.4\times10^{-1}$ & $1.7\times10^{-1}$& $1.6\times10^{-1}$ \\
 950 & $1.2\times10^{-1}$ & $1.2\times10^{-1}$ & $1.4\times10^{-1}$& $1.4\times10^{-1}$ \\
1000 & $1.1\times10^{-1}$ & $1.1\times10^{-1}$ & $1.3\times10^{-1}$& $1.2\times10^{-1}$ \\
1050 & $8.9\times10^{-2}$ & $9.4\times10^{-2}$ & $1.1\times10^{-1}$& $1.0\times10^{-1}$ \\
1100 & $6.7\times10^{-2}$ & $7.1\times10^{-2}$ & $8.5\times10^{-2}$& $8.2\times10^{-2}$ \\
1150 & $5.8\times10^{-2}$ & $5.9\times10^{-2}$ & $7.1\times10^{-2}$& $6.9\times10^{-2}$ \\
1200 & $4.6\times10^{-2}$ & $4.8\times10^{-2}$ & $5.8\times10^{-2}$& $5.7\times10^{-2}$ \\
1250 & $3.6\times10^{-2}$ & $3.6\times10^{-2}$ & $4.4\times10^{-2}$& $4.5\times10^{-2}$ \\
1300 & $2.6\times10^{-2}$ & $2.7\times10^{-2}$ & $3.3\times10^{-2}$& $3.4\times10^{-2}$ \\
1350 & $1.8\times10^{-2}$ & $1.9\times10^{-2}$ & $2.3\times10^{-2}$& $2.4\times10^{-2}$ \\
1400 & $1.4\times10^{-2}$ & $1.4\times10^{-2}$ & $1.6\times10^{-2}$& $1.8\times10^{-2}$ \\
\end{tabular}
\end{ruledtabular}
}
\label{tab:limit}
\end{table}
\endgroup

The obtained 95\% confidence level (C.L.) 
limits using the $W'$, $Z'$, RS graviton,
and $q^*$ signal resonance shapes are shown in
Fig.~\ref{fig:limit_theory_vs_mass} and Table~\ref{tab:limit}
as a function of the new particle mass.
Also shown in Fig.~\ref{fig:limit_theory_vs_mass}
are the theoretical predictions for the various models.
For the $W'$, $Z'$, $q^*$, and RS graviton,
the mass exclusion is determined by comparing the
limits obtained with their respective signal shapes and 
$\sigma^{sig}$ predictions.
For the axigluon, flavor-universal coloron, and $E_6$ diquark, we
compare their
$\sigma^{sig}$ predictions to the limits obtained with the $q^*$
signal shapes; these particles do not decay into the mode
containing a gluon, so their signal shape would be narrower
than that of $q^*$, and thus the mass exclusions obtained
with the $q^*$ signal shapes are conservative.
For $\rho_{T8}$, we compare its $\sigma^{sig}$ 
predictions with the limits obtained using the RS graviton
signal shapes.
The $\rho_{T8}$ and RS graviton decay
channels are similar; the branching fraction to the $gg$ state is
higher for the RS gravitons than for $\rho_{T8}$, so comparing
the limits obtained using the RS graviton signal shapes to the
$\rho_{T8}$'s $\sigma^{sig}$ predictions yields a conservative
mass exclusion.

The mass exclusion regions obtained in this search are
$260<m<870\mbox{~GeV}/c^2$ for $q^*$, 
$260<m<1250\mbox{~GeV}/c^2$ for the axigluon and flavor-universal coloron, 
$290<m<630\mbox{~GeV}/c^2$ for the $E_6$ diquark,
$260<m<1100\mbox{~GeV}/c^2$ for $\rho_{T8}$, 
$280<m<840\mbox{~GeV}/c^2$ for $W'$, and
$320<m<740\mbox{~GeV}/c^2$ for $Z'$.
For the RS graviton, this search did not exclude any mass region.
As the mass exclusions from the previous dijet mass resonance
searches~\cite{cdf1a,cdf1b,dzero} are
$m<775$ GeV/$c^2$ for $q^*$,
$m<980$ GeV/$c^2$ for the axigluon and flavor-universal coloron,
$290<m<420$ GeV/$c^2$ for the $E_6$ diquark,
and $260<m<480$ GeV/$c^2$ for $\rho_{T8}$,
this search provides the most stringent lower mass limits for these
particles.
This search also extends the existing mass exclusions of
$300<m<800$~GeV/$c^2$ for $W'$ and $400<m<640$~GeV/$c^2$ for $Z'$ 
obtained in the previous dijet mass resonance search~\cite{dzero}.
However, the more restrictive mass exclusions of
$m<1000$ GeV/$c^2$ for $W'$ and $m<923$ GeV/$c^2$ for $Z'$ come from
the $W'\to e\nu$ search~\cite{dzero_enu} and $Z'\to e^+e^-$ 
search~\cite{cdf_ee}.

%
\section{\label{sec:conclusions}Conclusions}

In conclusion, we performed a search for new particles which decay into
dijets by measuring the dijet mass spectrum using 
$p\bar p$ collision data from 1.13 fb$^{-1}$ of integrated luminosity
collected with the CDF II detector.
The measured dijet mass spectrum is found to be consistent with NLO
pQCD predictions based on recent PDFs and does not show evidence of a
mass resonance from new particle production.
We set 95\% confidence level upper limits on new particle production
cross sections times the branching fraction
to dijets times the acceptance for both jets to have $|y|<1$.
We also determine the mass exclusions for the
excited quark, axigluon, flavor-universal coloron, $E_6$ diquark,
color-octet techni-$\rho$, $W'$, and $Z'$ for a specific
representative set of model parameters.
This search sets the most stringent lower mass limits
on the excited quark, the axigluon, the flavor-universal coloron,
the $E_6$ diquark, and the color-octet techni-$\rho$.

\begin{acknowledgments}
%
We thank the Fermilab staff and the technical staffs of the
participating institutions for their vital contributions. This work
was supported by the U.S. Department of Energy and National Science
Foundation; the Italian Istituto Nazionale di Fisica Nucleare; the
Ministry of Education, Culture, Sports, Science and Technology of
Japan; the Natural Sciences and Engineering Research Council of
Canada; the National Science Council of the Republic of China; the
Swiss National Science Foundation; the A.P. Sloan Foundation; the
Bundesministerium f\"ur Bildung und Forschung, Germany; the Korean
Science and Engineering Foundation and the Korean Research Foundation;
the Science and Technology Facilities Council and the Royal Society,
UK; the Institut National de Physique Nucleaire et Physique des
Particules/CNRS; the Russian Foundation for Basic Research; the
Ministerio de Ciencia e Innovaci\'{o}n, and Programa
Consolider-Ingenio 2010, Spain; the Slovak R\&D Agency; and the
Academy of Finland. 
\end{acknowledgments}

\bibliography{mjj_search_paper_draft}

\begin{thebibliography}{43}
\bibitem[{coo()}]{coordinate}
\bibinfo{note}{We use a cylindrical coordinate system with the $z$ coordinate
  along the proton beam direction, the polar angle $\theta$, and the origin at
  the center of the detector. The rapidity, pseudorapidity, transverse energy,
  and transverse momentum are defined as
  $y=\frac{1}{2}\ln\left[(E+p_z)/(E-p_z)\right]$, $\eta=-\ln\tan(\theta/2)$,
  $E_{T}=E\sin(\theta)$, and $p_{T}=p\sin(\theta)$, respectively.}
\bibitem[{cal()}]{caltow}
\bibinfo{note}{The four-vector for each calorimeter tower is computed as a sum
  of vectors for the electromagnetic and hadronic compartments of the tower;
  the vector for each compartment is defined by assigning a massless vector
  with magnitude equal to the deposited energy and with direction from the
  primary event vertex to the center of each compartment.}
\bibitem[{mis()}]{missingEt}
\bibinfo{note}{The missing $E_T$ is defined by $\met = |\mathbin{\vec{E}\mkern
  - 11mu/_T}|$, $\mathbin{\vec{E}\mkern - 11mu/_T} = -\sum_{i} E_T^i {\bf
  \hat{n}_i}$, where ${\bf \hat{n}_i}$ is a unit vector perpendicular to the
  beam axis and pointing at the i$^{th}$ calorimeter tower. The sum $E_T$ is
  defined by $\sum{E_T} = \sum_{i} E_T^i$. Both sums are over all the
  calorimeter towers with $E_T>100$ MeV.}
\end{thebibliography}

\end{document}